\documentclass[runningheads,a4paper]{llncs}
\usepackage{amssymb}
\usepackage{graphicx}
 \usepackage{amsmath}
 \usepackage{subcaption}
\usepackage{multirow}
 \usepackage{paralist}
 \usepackage{adjustbox}
 \usepackage[misc]{ifsym}
 
\begin{document}

\mainmatter  

\title{Deep Generative Model-Driven Multimodal Prostate Segmentation in Radiotherapy}
 
\titlerunning{Deep Generative Model-Driven Organ Boundary Detection}

%
\author{Kibrom Berihu Girum\inst{1,2 (\textrm{\Letter})} \and Gilles Cr\'ehange\inst{1,2} \and Raabid Hussain\inst{1} \and Paul Michael Walker\inst{1,3} \and Alain Lalande\inst{1,3}}
\authorrunning{K.B. Girum et al.}

\institute{$^{1}$ImViA, Universit\'e de Bourgogne Franche-Comt\'e, Dijon, France\\
	\textsuperscript{\textrm{\Letter}}\email{kibrom-berihu\_girum@etu.u-bourgogne.fr} \\
	$^{2}$Department of Radiation Oncology, CGFL, Dijon, France\\
$^{3}$Depratment of Medical Imaging, University Hospital of Dijon, Dijon, France  \\ 
}
%
%

\toctitle{Organ Boundary Detection}
\tocauthor{Deep Generative Model}
\maketitle

\begin{abstract} 
\label{sec:abstract}

Deep learning has shown unprecedented success in a variety of applications, such as computer vision and medical image analysis. However, there is still potential to improve segmentation in multimodal images by embedding prior knowledge via learning-based shape modeling and registration to learn the modality invariant anatomical structure of organs. For example, in radiotherapy automatic prostate segmentation is essential in prostate cancer diagnosis, therapy, and post-therapy assessment from T2-weighted MR or CT images. In this paper, we present a fully automatic deep generative model-driven multimodal prostate segmentation method using convolutional neural network (DGMNet). The novelty of our method comes with its embedded generative neural network for learning-based shape modeling and its ability to adapt for different imaging modalities via learning-based registration. The proposed method includes a multi-task learning framework that combines a convolutional feature extraction and an embedded regression and classification based shape modeling. This enables the network to predict the deformable shape of an organ. We show that generative neural network-based shape modeling trained on a reliable contrast imaging modality (such as MRI) can be directly applied to low contrast imaging modality (such as CT) to achieve accurate prostate segmentation. The method was evaluated on MRI and CT datasets acquired from different clinical centers with large variations in contrast and scanning protocols. Experimental results reveal that our method can be used to automatically and accurately segment the prostate gland in different imaging modalities. 
 
\end{abstract}

\section{Introduction}
\label{sec:introduction}

\begin{figure}[h]
	\centering
	\includegraphics[width=.9\textwidth]{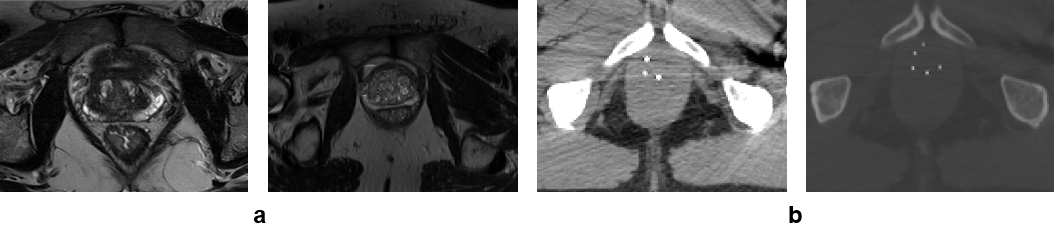}
	\caption{\small {Prostate image examples showing image contrast variations in: (a) T2-weighted MRI, and (b) CT images with seeds from low-dose-rate brachytherapy.}}
	\label{fig:prostate_images}
\end{figure}

Automatic segmentation of anatomical structures in medical images has various medical applications. For example, in radiotherapy prostate segmentation is essential in the diagnosis, therapy, and post-therapy analysis of prostate cancer. It is critical in selecting patients for a specific treatment, to guide source delivery and in computing dose distribution \cite{multiscale,PPB}. T2-weighted MRI is the modality of choice for prostate segmentation. However, CT and US are also routinely used because: 
\begin{inparaenum}[1)]
	\item CT image is used to calculate the dose distribution due to its characteristics of relating the density of tissues with the voxel intensity, and 
	\item US imaging is suitable for real-time image guided radiotherapy.
\end{inparaenum}
Despite the need for accurate segmentation of the prostate in radiotherapy, manual segmentation is subjective to inter and intra-observer variabilities, time-consuming, and depends on the experience of the physician. Automatic and reliable segmentation of the prostate on these images is thus an important but difficult task due to the inhomogeneous and inconsistent contrast of prostate boundary and large shape variations. This is particularly complicated on CT images because of the inherent low-contrast imaging characteristics of CT for soft tissues (such as prostate boundary) as can be seen from Fig. \ref{fig:prostate_images} (b).  
 
Recently, organ boundary detection through modeling and incorporating the organ shape as prior information has been successfully used for automatic and reliable anatomical structure segmentation (such as  prostate \cite{multiscale}, brain \cite{brian}, and heart \cite{heart}). The prior prostate shape has been modeled using principal component analysis from labeled prostate CT scans. The modeled shape was then used to guide the segmentation of prostate gland on CT images \cite{multiscale}. A deep learning approach followed by a multi-atlas based feature extraction has also been proposed \cite{multi-atlas}. Distinctive curve guided fully convolutional neural network has also been employed for the pelvic organ segmentation on CT images \cite{distinictivecurve}. Kazemifar et al. \cite{organatrisk} used convolutional networks (U-net architecture \cite{Unet}) to segment both the prostate and organs at risk in male pelvic CT images. Guo et al. \cite{Deepfeaturelearning} have also used deep features and sparse patch matching approach to segment the prostate on MR images. Although atlas and shape-prior based methods demonstrated promising performance, they might not be generic. This lack of generalization ability is due to the possibility of statistical shape or atlas of an organ being different for a new patient. It then requires a robust modeling and registration algorithm. However, robust feature extraction is still a challenging task to obtain an optimal shape model. Indeed in medical images, image contrast, organ shape, acquisition protocol and deformable characteristics of an organ can vary widely.
  
Deep convolutional neural networks (CNN) have shown promising performances in various medical applications \cite{deeplearning}. For example, U-net architecture has been often used for medical image segmentations \cite{Unet}. Adversarial neural network has also been proven to improve medical image segmentation (e.g. for liver \cite{liver}). Thus, our hypothesis is that by combining CNN-based feature extraction and learning-based anatomical structure modeling (through generative neural network) from reliable contrast images (such as T2-weighted MRI for soft tissues), we can predict accurately an organ boundary in low-contrast imaging modalities (e.g. prostate segmentation on CT).

In this paper, we present a new deep generative model-driven anatomical structure segmentation (named DGMNet), specifically designed for multimodal (CT and MR) prostate segmentation. The proposed method employs a convolutional feature extraction with an embedded generative CNN \cite{Unet,senetwork}. The generative CNN is designed for learning-based modeling of prior organ shape from MRI and applied to low-contrast CT images. It also involves a learning-based registration with a given raw input image. Experimental results on MRI and CT datasets reveal that our method can fully automatically segment the prostate robustly and accurately regardless of the difference in  contrast, size, and imaging modality.  

\section{Methodology} 
\label{sec: proposed method}
We aim at detecting the boundary of the prostate volume in a given 3D raw input image $I$  of size $W \times H \times C$. Here $W$, $H$, and $C$ are width, height and depth of the image, respectively. We use a deep CNN which outputs a label map of size  $W \times H \times C$ whose voxels $v=(x,y,z)$ contain a label 1 for prostate volume and 0 otherwise. This is done by combining a predicted label mask from a decoder and a predicted shape from an embedded shape-model generator (see Fig. \ref{fig:model} (a)).  

\subsection{Network architecture}
\label{subsec: Networkarchitecture}

\begin{figure}[h]
	\centering
	 \includegraphics[width=0.88\linewidth]{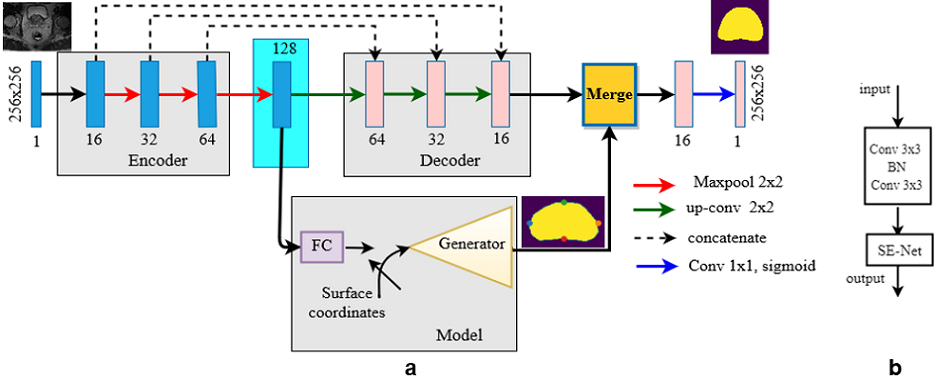}
	  \caption{ \small { Proposed architecture: (a) The overall framework (DGMNet), and (b) the schema of a single block in the encoder and decoder.}}
	   \label{fig:model}
\end{figure}

The network architecture is illustrated in Fig. \ref{fig:model}. It consists of feature extraction, generative shape modeling, and feature map upsampling. The feature extraction (encoder) resides on a convolutional neural network \cite{senetwork}. It consists of a repeated application of 3x3 convolution, Rectified Linear Unit (ReLU) activation, batch normalization, followed by squeeze-and-excitation network (SE-Net) \cite{senetwork}(see Fig. \ref{fig:model} (b)), and a 2x2 max pooling operation with stride 2 for downsampling. From the extracted feature maps, two paths named as model and decoder path are applied. We also used dropout regularization at the bottleneck layer to have a better generalization by reducing over-fitting during the training. The decoder path is composed of a 2x2 up-convolution and concatenation layer, followed by the same block as the encoder (Fig. \ref{fig:model} (b)). Generative path (i.e. Model in Fig. \ref{fig:model} (a)) is composed of average max pooling followed by fully connected layers (FC). The output of these FC layers (corresponding to surface boundary coordinates) are feed to the generative model where it generates the shape of a given organ (in our case, prostate gland). It consists of a projection and a reshape block followed by repeated Leaky ReLU activation (except the last activation which was sigmoid), batch normalization, and up-convolution (similar to the one proposed in \cite{generative}). The model-generator and decoder outputs are merged using addition and further a convolutional block is applied. The output layer is a 1x1 convolutional layer with sigmoid activation function. It is worthy to mention here that the proposed network involves only 1.5 million trainable parameters while the U-net architecture has 31.024 million \cite{Unet}.

\subsubsection{Generator:}
\label{subsec:generator}

We formulate the model generator to predict the prostate volume given a few sampled prostate surface boundary coordinates, $I^{u} = (x^{u}, y^{u}, z^{u})$. To this end, the voxel depth ($z$, for a given slice $u$) is taken as classification task (0 or 1)  and the remaining ($x$ and $y$) as regression task. Given the surface boundary coordinates, $I^{u} = (x^{u}, y^{u}, z^{u})$, it is trained to predict a labeled model  $ W \times H \times C $, in which the prostate volume is 1 and 0 otherwise, i.e. $ G(I^{u}) \mapsto W \times H \times C$. We automatically extracted four surface boundary landmark coordinates (left, right, top, and bottom) per-slice from the given labeled ground truth (from MRI) and repeated over the whole volume of the prostate ($I^u$).
 
\subsection{Loss function}
\label{subsec: lossfunction}

To train the proposed network, we define a multi-task loss function as a combined weighted sum loss:
\begin{equation} 
L_{total} = L_{mask} + \lambda L_{clsLnd} \;  ,
\label{eq:one}
\end{equation} 

in which the segmentation (final mask) loss, $L_{mask}$, is calculated as a combination of Dice and cross entropy loss ($L_{mask}= L_{dice} + L_{CE}$). 

Given ground truth surface voxel coordinates $I^{u} = (x^{u}, y^{u}, z^{u})$, where $z^{u} = 1$, and predicted values, $I^{v} = (x^{v}, y^{v}, z^{v}=p)$, the joint classification and regression loss can be calculated as:

\begin{equation}
L_{clsLnd} (I^{u}, I^{v}) = L_{cls} (p, z^{u}) + [z^{u} = 1]L_{lnd} (t^{u}, t^{v}) \;  ,
\label{eq:two}
\end{equation}

in which the classification loss, $L_{cls}(p, z^{u})$, is the cross entropy loss. The likelihood of a given raw input image slice ($I ^u$), being part of the organ is $p$. The ground-truth label, $z^{u}$, is 1 if the image slice consists of the prostate, and is 0 otherwise. The second loss, $L_{lnd} (I^{u}, I^{v})$, is thus defined over the surface landmarks where the ground truth is 1 and 0 otherwise. For positive ground truth (i.e. $z^{u} = 1$), we use smooth $L1$ loss between corresponding voxels, which is considered as robust loss to outliers \cite{fastrcnn}, as:

\begin{equation}
L_{lnd}(t^{u}, t^{v})  =  \sum_{i\in \{x, y, z=1\} }^{} smooth_{L{1}} (t^{u}_{i} - t^{v}_{i}) \;  ,
\label{eq:three}
\end{equation} 

in which 

\begin{equation}
smooth_{L1}(\Delta t)=\left\{
\begin{array}{@{}ll@{}}
0.5(\Delta t)^2   & \text{if}\ |\Delta t|<1 \\

|\Delta t|-0.5  \hspace{1cm}  & \text{otherwise}  \;  ,
\end{array}\right.
\label{eq:four}
\end{equation}

where $t^{u} = ({x^{u}, y^{u}})$ and $t^{v} = ({x^{v}, y^{v}})$ are the ground truth and predicted surface boundary coordinates, respectively, for given $z^{u} = 1$. The hyper-parameter $\lambda$ controls the losses contributed from the segmentation and surface boundary coordinates.   

\section{Experimental setup and results}
\label{sec:experimentalsetupandresults}

\subsection{Datasets}
\label{subsec:dataset}

The proposed method was trained and evaluated on T2-weighted MRI and CT prostate images with vast variability in organ size, shape, scanning protocol, and from multiple clinical centers. Firstly, it was trained and evaluated on 60 T2-weighted MR exams. These datasets were acquired with an in-plane resolution ranging from $0.312$x$0.312$ $mm^2$ to  $0.676$x$0.676$ $mm^2$ with a slice thickness between 1.250 mm and 2.722 mm. Similarly, we also trained and evaluated on 40 CT patient datasets (who underwent permanent prostate brachytherapy with $^{125}I$ for localized prostate cancer treatment). These CT exams were acquired from two clinical centers. The in-plane resolution  of these CT data varies from $0.4$x$0.4$ $mm^2$ to $0.58$x$0.58$ $mm^2$ with a slice thickness between 1.5 mm and 2.5 mm (helical mode, 120 kVp, 172 mm FOV, and 440 mAs/slice). We looked for conversions of the datasets into the same voxel size of $0.5$x$0.5$x$1.25$ $mm^3$ and $0.7$x$0.7$x$1.25$ $mm^3$ for CT and MRI respectively. The prostate was manually delineated by experienced radiologists.

\subsubsection{Pre-processing: } The input images (MR and CT) were pre-processed  by zero-centering the intensity values and normalizing them by the standard deviation of all images before feeding to the network. All images were also center cropped and resized to have an image resolution of 256x256.  

\subsubsection{Training and testing details:} We trained the system by minimizing the loss $L_{total}$ (equation \ref{eq:one}). The proposed system is trained as follows:
\begin{inparaenum}[1)] 
	\item Firstly, we train the generative model with the inputs from a few sampled surface boundary landmarks of the prostate volume, specifically from only T2-weighted MRI labels. We used a binary cross entropy loss for training. We conducted five-fold cross validation experiment. 
	\item Secondly, the whole system is trained except the generator in which it only predicts the model shape given the predicted surface coordinate values from FC network.
\end{inparaenum}
We use a batch size of 10 images for both MRI and CT. It is important to mention here that we feed the network with 2D instead of 3D as we have small datasets. Then, predicted image labels are stacked to create a 3D volume. The model is trained using Adam optimizer with a learning rate of $0.001 $. The whole ensembled architecture (except the generator) was trained from scratch considering 10 patients (25\% of the datasets which were selected randomly) for validation. 

\subsubsection{Ablation study:}
We conducted ablation experiments (Table \ref{tableall}) to investigate the effect of individual components in the proposed network. All ablation experiments were done on CT images under similar settings:
\begin{inparaenum}[1)] 
	\item Unet architecture; 
	\item ResUnet (residual-based Unet);
	\item SE-ResUnet (residual block with squeeze-and-excitation network based Unet \cite{senetwork});  
	\item SE-Unet (the proposed method without the generator and the FC network (Fig. \ref{fig:model} (b))). 
\end{inparaenum} 

\subsubsection{Evaluation metrics:}All experiments were evaluated using Dice Similarity Coefficient (DSC), Sensitivity (Sen), positive predicted value (PPV), and average surface distance (ASD) (in mm) \cite{promis2012}.

\subsection{Experimental results}
\begin{figure}[h]
	\centering
	\includegraphics[width=0.8\textwidth]{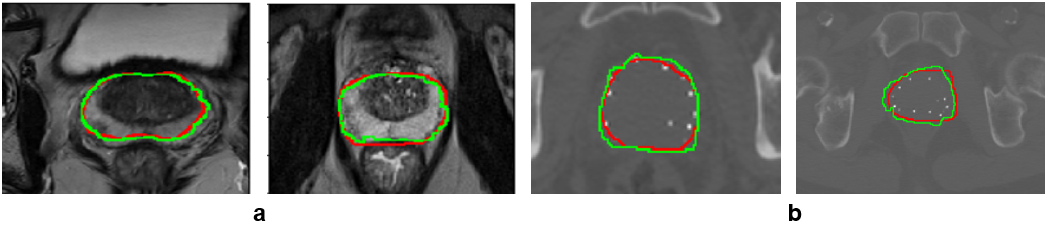}
	\caption{ \small {Qualitative evaluation of prostate segmentation on 2D: (a) T2-weighted MRI, and (b) CT images with seeds from low-dose-rate brachytherapy. The ground truth labels are shown in red and segmentation results in green.}}
	\label{fig:segmentationresult}
\end{figure}

The generator was trained using a five-fold cross validation method using T2-weighted MRI. It was then kept as a shape predictor by freezing its weights during training  of the proposed method. As this is an intermediate output of the method, it can be considered as a region proposal (or as instantaneous shape generator) to be further refined by merging with the encoder-decoder output. Indeed, the model-generator can learn from good contrast images (MRI) and used directly (transfer without fine tuning by freezing) for low contrast images (CT), while the encoder-decoder extracts additional features. As one can see from the qualitative prostate segmentation results in Fig. \ref{fig:segmentationresult}, the proposed method can segment accurately the prostate on both T2-weighted MR and CT images. 

In almost all evaluation metrics (with and without the generator, Table \ref{tableall}), the proposed method with the shape model generator outperforms the state of the art methods. Since the implanted radioactive seeds were not uniformly placed over the volume of the prostate gland, it was observed to influence the segmentation quality (particularly the state of the art methods). However, they might perform better on CT images without the implanted radioactive seeds. Combining CNN-based extracted features with prior shape knowledge of the organ can improve time, reproducibility, and accuracy in fully automatic segmentation of the prostate in radiotherapy.

\begin{table}[h!]
	\centering		 		
	\caption{Quantitative segmentation results. Values are expressed as mean $\pm$ std.} 	
	\label{tableall}
	\begin{adjustbox}{width=\textwidth}
		\small
		\begin{tabular}
			{c c c c c c  c c}
			\hline
			Data & Method & DSC & Sen & ASD &  PPV \\  
			\hline
			\multirow{5}{*} {CT} \hspace*{0.25cm}	&  Unet & 0.83 $\pm$ 0.04 & 0.76 $\pm$ 0.0.08 & 0.16 $\pm$ 0.08 &  0.93 $\pm$ 0.03  \\ 
			&  ResUnet & 0.82 $\pm$ 0.03  & 0.73 $\pm$ 0.03 & 0.16$\pm$ 0.10 &  0.93 $\pm$ 0.03 \\
			&  SE-ResUnet & 0.84 $\pm$ 0.03 & 0.88 $\pm$ 0.05 &  0.84 $\pm$ 0.53 &  0.82 $\pm$ 0.06 \\
			&  SE-Unet & 0.85 $\pm$ 0.03 & 0.78 $\pm$ 0.07 &  0.17 $\pm$ 0.15 &  0.93 $\pm$ 0.04 \\
			& \textbf{DGMNet} & 0.89 $\pm$ 0.02  & 0.92 $\pm$ 0.03 & 0.28 $\pm$ 0.09 &  0.87 $\pm$ 0.03\\ 
			
			\hline
			\multirow{1}{*} {MRI} & 
			\textbf{DGMNet} & 0.93 $\pm$ 0.12 & 0.92 $\pm$ 0.15 & 0.11 $\pm$ 0.22 & 0.96 $\pm$ 0.07\\
		\end{tabular}
	\end{adjustbox}
\end{table}

\section{Conclusion}
\label{sec:conclusion}

In this paper we proposed DGMNet, a new  CNN approach for feature-model learning based anatomical structure segmentation. It is an encoder-decoder architecture and an embedded deep  generative neural network based model-generator that enables training on limited data. The model-generator is used for embedding prior shape knowledge via learning based shape modeling and registration from high contrast images (such as MRI) and directly applied (by freezing) to low contrast images (such as CT). Further, we demonstrated that combining shape-model with a CNN-based feature extraction improves segmentation accuracy. We extensively evaluated models trained with and without prior shape generator on CT images with different metrics to verify the effect of the embedded shape generator. Experimental results, on MR and CT datasets, reveal that this method can be used to fully-automatically segment prostate gland in different imaging modalities. In the future, we plan on generalizing the proposed method to other modalities such as US images (for intra-operative radiotherapy) as well as to other organs (such as rectum, brain, and heart). In the case of US images, we shall propose to train the model-generator from MRI with an endorectal coil to consider the deformation characteristics of the prostate gland from the coil.

\end{document}